\documentclass[12pt]{article}

\usepackage[title]{appendix}

\author{A.S.~Yurkov \\ 
Omsk, Russia, 644076, e-mail:fitec@mail.ru }

\title{Elastic Boundary Conditions in the Theory with Second Gradients in the Thermodynamic Potential}

\begin{document}

\maketitle

\begin{abstract}
When describing elastic deformations of a body sometimes  it is worth  to take in account elastic spatial dispersion. If spatial dispersion is weak, as usually happens, then it can be reduced to dependence of thermodynamic potential on strain gradients.  Such a dependence may be worth in association with  small body size which imply large gradients.  Besides,  the inclusion of this  dispersion leads to physical phenomena absent without it. An example of the latter is flexoelectricity.  Remarkable fact is that while the derivation of differential equations of elastic equilibrium can be made by ordinary means in this case,  the derivation of  boundary conditions for them  is less trivial task. This is due to the fact that strain gradients should be represented in terms of second gradients of independently varied elastic displacements. Detailed consideration of this problem is the subject of this paper.
\end{abstract}

\section{Introduction}

The classical theory of elasticity is based on the thermodynamic potential depended on the elastic strain only. However, in some cases it is important to consider  dependence of the thermodynamic potential on  the gradients of  elastic strain.

There are two main reasons why  consideration of such  a thermodynamic potential  may be important. First, elastic  body can  be of small size, small sizes imply large gradients. This becomes especially important due to rapid development  of nanotechnology. Secondly,  such a dependence  may lead to  physical phenomena absent without it. An example of such  phenomenon is the flexoelectric effect (see \cite{bib:tagyud13,bib:zubko13} and references therein).

The basic problem of the theory of elasticity is to find  static equilibrium of a body. Generally speaking dynamic problems are also important, but we restrict ourselves to the static case. Generalization of the considered  theory to the dynamic case is  absolutely straightforward.

Even if thermodynamic potential depends on strain gradients, it remains  quasi-local potential (it can be represented as integral of the corresponding density).  Thus, equilibrium is described by differential equations in partial derivatives. This equations should be appended with  boundary conditions on the surface of the body. Both  differential equations and boundary conditions are determined by the conditions of thermodynamic potential minimum in framework of the calculus of variations. 

It is important to note that in the considered case the derivation of differential equations of elastic equilibrium does not differ much from the case of the classical theory, it can be made by completely standard methods. On the contrary, the correct derivation of the boundary conditions for these equations is non-trivial task. Recent studies on the subject  have been done with regard to the description of the flexoelectric effect, but they are applicable in all cases when the thermodynamic potential depends on  strain gradients. 
 
Essence of  matter leading to the fact that in this case the problem of boundary conditions is non-trivial   is as follows. Dependence of the thermodynamic potential of the strain gradient actually means its dependence on the second spatial derivatives of the elastic displacements $u_i\,$. When calculating the variation of the thermodynamic potential one should  express it in terms of  the independent variate. As usual, this is done using integration by parts. However, in the considered case  surface integrals, which arises due to such an integration,  contained  $\delta u_{i,j}$ not just $\delta u_i\,$. Hereinafter  index separated by comma denotes spatial derivative in respective coordinate. Sometimes such derivative is also denoted as $\partial_i\,$.  Values $\delta u_{i,j}$ are not independent variates. Thus, additional analysis and mathematical constructions are required. 

Apparently a  first solution to this problem has been  proposed by Toupin \cite{bib:Toupin62,bib:Toupin64} many years ago. This solution is based on the
 representation of the gradient $\partial_i$ as a sum of  ``normal gradient'' $n_iD$ and ``tangential gradient'' $D_i\,$, where $n_i$ is unit vector normal to the surface. Operator $D$ has been defined as $n_i\partial_i\,$. Hereinafter the  Einstein summation rule over repeated indices is implied. Note that formal definition of $D_i$ has not been proposed. Certainly, if $D_i$ is applied to a function, defined in whole 3-dimensional space (at least in the body bulk), then $D_i$ can be well-defined as difference between  $\partial_i$ and $n_iD\,$, no problem appears in this way.  However, Toupin's boundary conditions contains not only such mathematical objects, but also $B_{ij}=D_in_j\,$. Vector $n_i$ is defined only on the surface  and hence neither $\partial_i$ nor $D$ can not be applied to it.  Thus,  mathematical sense of 3-dimensional tensor $B_{ij}$ has been remained unclear. Of course, this construction  can be provided with some intuitive sense in  Cartesian coordinate system, if a couple  of  axes are tangential to the surface at  given point. But anyway it is not flawless mathematical definition, because tangential plane has the only point common with the curved surface. Formally,  vector $n_i$ is \emph{not} defined in some, at least infinitesimal, region of this  tangential plane. Naturally,   the derivative can be defined only if differentiated expression is defined in  some region.

In \cite{bib:Toupin62,bib:Toupin64} tensor $B_{ij}$  has been referred as ``second fundamental form'' of the body surface. Note that second fundamental form of the surface, at least within conventional definitions, is defined in 2-dimensional  space tangential to the surface,  rather then in whole 3-dimensional space. Thus,  relation between  second fundamental form and 3-dimensional tensor $B_{ij}$ remains unclear. At least additional geometric constructions and definitions are required here to make it  meaningful.

In subsequent years, Toupin's construction has been used to derive boundary conditions in many papers (see \cite{bib:Mindlin_Eshel_68,bib:Germain73,bib:shul10} for instance). However, mathematically flawless definition of  $D_in_j$ has not been given. Moreover, in these papers it was not even mentioned the relation between this tensor and  the second fundamental form of the boundary surface.

In the framework of continual theory of flexoelectricity an attempt to derive the mechanical boundary conditions   was made also in the paper \cite{bib:inden81}. The authors of this paper quite correctly pointed out that the mechanical boundary conditions can be obtained from the surface integral arising when  the thermodynamic potential is varied in the elastic displacements.  It was also correctly noticed  that since the thermodynamic potential depends on strain gradients in this case, the  differential equations are of  the fourth order in contrary to  the standard theory of elasticity, where they are of the second order. Therefore, the number of the boundary conditions should be grater then such a number  in the classical theory of elasticity.

Nevertheless  mechanical boundary conditions  written in \cite{bib:inden81}    can not be considered as correct ones. Authors do not give  detailed derivation of them, so it is not possible to  specify  where a mistake done. However, erroneousness of these boundary conditions  is clear already from the fact that they do not contain  the  vector normal to the boundary surface.  This means,  among others, that  it is impossible to passage to the limit  of the classical theory of elasticity where boundary conditions contain such a vector.

In addition it is obvious that the authors of  \cite{bib:inden81} did not take in account the curvature of the boundary surface. The surface of the finite-size body can not be flat elsewhere,  further it is shown that in the considered case the  surface curvature plays an important role.

Number of  articles \cite{bib:MSS06,bib:SMS07,bib:MSC08} have been published  also  where for the case of presence of flexoelectricity mechanical boundary conditions on the free surface of a body  were declared in the form   $\sigma_{ij}n_j=0$  which is analogous to the form  of the boundary conditions in the classical theory of elasticity. Here $\sigma_{ij}$ is tensor of  non-classical ``physical stress'' which enters into differential equations of equilibrium in the form $\sigma_{ij,j}=0$.  Note that  this  boundary conditions were simple declared without any derivation. Note also that, as mentioned above,  number of boundary conditions should be greater at least.

Mathematically strict  way to derive the  mechanical boundary conditions in the presence of flexoelectricity was independently proposed  in the paper \cite{bib:yurk11}. In this paper, in contrast  earlier works,  projector to the body surface, needed to decompose $\delta u_{i,j}$ to normal and tangential parts, was constructed  by means of  two-dimensional curvilinear coordinate system on the body surface. This two-dimensional coordinate system  was introduced  additionally to common three-dimensional Cartesian coordinate system in the space. Such mathematical construction does not lead to ill-defined mathematical objects. Note that although in this paper  the boundary conditions has been derived in framework of the theory of flexoelectricity,  in fact they can be applied to  all cases where  thermodynamic potential depends on  strain gradients.

Latter  yet another approach to the derivation of such  boundary conditions  was proposed in the paper  \cite{bib:yurk13}. This approach is  based on a mathematical tool which is almost identical to the one used in  general theory of relativity. The only difference is  that, since the space is flat, the Riemann tensor is equal to  zero and  covariant derivatives commute. 

Note that  for solution of certain  problems the curvilinear coordinates\footnote{Note that  index distinguishing curvilinear coordinates  conventionally written as superscript instead of subscript.} $x^{\alpha}$ are more convenient then Cartesian ones, especially  if one choose a coordinate system, in which the equation of the body surface is  $x^3=x^3_S\,$, where $x^3_S$ is some constant. In such  coordinate system transformation of surface integrals needed for derivation of  discussed  boundary conditions is performed directly without any mathematical tricks. That is exactly what was used in \cite{bib:yurk13}.

Detailed derivation of the boundary conditions in curvilinear coordinates has not been given in \cite{bib:yurk13}, there  was only a brief note  about  the method of derivation. It  was also omitted  the proof  that this boundary conditions are completely equivalent to the boundary conditions derived in   \cite{bib:yurk11}. This equivalence was only mentioned without any evidence. Consideration was too brief  in \cite{bib:yurk13}.

Thus, detailed description of the problem of mechanical boundary conditions for a case where the thermodynamic potential depends on strain gradients is desirable. It is presented here and in somewhat more general form then in \cite{bib:yurk11,bib:yurk13}.

%\newpage

\section{General description of the mechanical equilibrium of the elastic body}

In the most general case, the mechanical equilibrium of an elastic body can be found in framework of calculus of variations by minimization of thermodynamic potential  $H$. The minimization should be performed with respect to variations of the elastic displacements $u_i$.  

Thermodynamic potential  also depends on variables other than $u_i$. Particularly,  in the theory of   flexoelectricity it also depends on  electric polarization at least. However, here we consider only the mechanical aspects of the theory, so that electric polarization and all other variables, which differ from $u_i\,$, will be considered as given.

In quasi-local case the thermodynamic potential of the body  is the sum of bulk part $H_B$ and surface part $H_S$. The bulk part is volume integral  of the bulk density of the thermodynamic potential:
\begin{equation}
H_B=\int {\cal H}_BdV \, .
\end{equation}
Analogously the surface part is an integral over the body surface of the surface density:
\begin{equation}
H_S=\oint {\cal H}_SdS \, .
\end{equation}

Note that the decomposition of the thermodynamic potential to volume and surface parts can be non-unique. If ${\cal H}_B$ contains a term which is divergence then the volume integral of this term can be converted to a surface integral. The reverse transformation is possible also if ${\cal H}_S$  contains a suitable term.

As indicated above to  find the body equilibrium it is necessary to calculate the variation of the thermodynamic potential and equate it to zero. The result of this calculation depends essentially  on the type of the functional dependence of  ${\cal H}_B$ and ${\cal H}_S$ on $u_i$.

In the classical theory of elasticity ${\cal H}_B$   depends only on the strain tensor $u_{ij}$ which is the symmetric part of the distortion tensor:
\begin{equation}
u_{ij}=\frac{1}{2}(u_{i,j} + u_{j,i}) \, .
\end{equation}
In the case considered here  ${\cal H}_B$ depends on $u_{ij}$ and also depends on $u_{ij,k}$.  Note that in the general reasoning it is quite possible to assume that ${\cal H}_B$  depends on $u_{i,j}$ and $u_{i,j,k}$ directly.  Real dependence only on $u_{ij}$ and  $u_{ij,k}$ is automatically ensured due to the specific form of this dependence in this way, in fact it is ensured by symmetry properties of the material tensors. Keeping this in mind further we use unsymmetrical distortion tensor $u_{i,j}\,$. 

As for ${\cal H}_S$ in the  classical case it  depends  on $u_i$ only,  particularly this dependence describes interaction with  the  external forces  acting on the body surface. In the case considered here  one  should take into account also the dependence of ${\cal H}_S$  on $u_{i,j}$. Dependence ${\cal H}_S$ on $u_{i,j}$ is of the same order of smallness as the dependence ${\cal H_B}$  on $u_{i,j,k}$. This is clear already from the fact that volume integral of ${\cal H}_B$ being a divergence can be transformed to surface integral, order of derivatives is reduced at this transformation.

For such ${\cal H}_B$ and ${\cal H}_S$ the variation $\delta H$ can be represented in terms of following  quantities:
\begin{equation}
T_{ij} = \frac{\partial {\cal H}_B}{\partial u_{i,j}} \, ,
\end{equation}
\begin{equation}
\Theta_{ijk} = \frac{\partial {\cal H}_B}{\partial u_{i,j,k}} \, ,
\end{equation}
\begin{equation}
F_i=\frac{\partial {\cal H}_S}{\partial u_i} \, ,
\end{equation}
\begin{equation}
\Sigma_{ij}=\frac{\partial {\cal H}_S}{\partial u_{i,j}} \, .
\end{equation}
In this notation following equations can be written immediately:
\begin{equation}
\label{eq:dhb}
\delta H_B = \int T_{ij}\delta u_{i,j}dV + \int \Theta_{ijk}\delta u_{i,j,k}dV \, ,
\end{equation}
\begin{equation}
\label{eq:deltaHs9}
\delta H_S = \oint F_i \delta u_i dS + \oint \Sigma_{ij} \delta u_{i,j}dS \, .
\end{equation}

Variations  $\delta u_i$ are independent but $\delta u_{i,j}$ and $\delta u_{i,j,k}$ are not. So that integrals in (\ref{eq:dhb}) should be transformed using integration by parts. This yields
\begin{equation}
\begin{array}{l}
\displaystyle
\delta H_B = -\int \left(T_{ij,j} - \Theta_{ijk,j,k}\right)\delta u_idV   \\ \\
\displaystyle 
+ \oint \left(T_{ij}-\Theta_{ijk,k}\right)\delta u_i n_j dS  + \oint\Theta_{ijk} \delta u_{i,j}n_k dS
\, ,
\end{array}
\end{equation}
where $n_i$ is  unit vector normal to the surface of the body. So that surface terms appear which are analogous to the terms in right hand side of  (\ref{eq:deltaHs9}).

Variations $\delta u_i$ are absolutely arbitrary. Particularly  $\delta u_i$ and $\delta u_{i,j}$ can be vanishing on the surface. In such a case  all surface integrals disappear,  remaining volume integral yields differential equations of equilibrium:
\begin{equation}
\label{eq:equil}
T_{ij,j} - \Theta_{ijk,j,k} = 0 \, .
\end{equation}
Note that if one defines a ``physical stress'' $\sigma_{ij}$ as follows 
\begin{equation}
\label{eq:sigdef}
\sigma_{ij}=T_{ij} - \Theta_{ijk,k}  \, 
\end{equation}
then   (\ref{eq:equil}) can be rewritten  in the  manner analogous to the case of the classical theory of elasticity:
\begin{equation}
\label{eq:sigequil}
\sigma_{ij,j}=0 \, .
\end{equation}

Boundary conditions for  equations  (\ref{eq:sigequil})  should be derived from the nullification of sum of  surface integrals for  $\delta u_i$ and $\delta u_{i,j}$ not vanishing on the surface. So that  boundary conditions are contained in the integral equation
\begin{equation}
\label{eq:intbound}
\oint \left( \sigma_{ij}n_j  + F_i\right)\delta u_i dS +
\oint \left(\Sigma_{ij} + \Theta_{ijk}n_k\right) \delta u_{i,j}dS = 0 \, ,
\end{equation} 
where we use (\ref{eq:sigdef}) to rewrite the first integral.

In the classical theory of elasticity $\Theta_{ijk}=0$, $\Sigma_{ij}=0$ so that there is no second integral in  (\ref{eq:intbound}) which contains $\delta u_{i,j}\,$. Therefore, in this theory there is no problem to derive the boundary conditions from (\ref{eq:intbound}) at ones. One need only equate to zero the expression in parentheses in the first integral. 

In the case when the thermodynamic potential depends on $u_{i,j,k}$ the situation becomes more complicated. Matter is that $\delta u_i$  and $\delta u_{i,j}$ can not be considered as completely independent. Therefore, to obtain the boundary conditions one should transform the second integral in left hand side of  (\ref{eq:intbound}). This transformation is non-trivial, in order to make it one needs a special mathematical tool which is considered in the following sections. Only after this transformation  mechanical boundary conditions can be obtained.

\section{Additional surface coordinate system and boundary conditions in Cartesian tensor components}\label{sec:cartbound}

As stated above, to derive the boundary conditions one must  transform the second integral in  the left hand side of (\ref{eq:intbound}).  One  method of  this transformation based on the use of additional surface coordinate system was proposed in \cite{bib:yurk11}. In this section we consider this method with some generalization relatively to the paper \cite{bib:yurk11}. This generalization is quite direct but  it can be useful sometimes.

Besides the usual Cartesian coordinate system in the space,  let be  an additional two-dimensional  coordinate system $x^{\alpha}$ on the surface of the body, $\alpha=1,2$ here. Generally the closed surface of a body is curved. Certainly body surface can be flat everywhere except  the sharp edges where flat parts of surface intersect. Such a situation can be considered as the limit of the body with smoothed edges for which  the curvature of the surface is present only on this smoothed edges. It is what further implied.

In this way a two-dimensional non-Euclidean geometry appears on the surface of a body. This is why the surface coordinate system should be described as curvilinear and one should distinguish corresponding tensor indices, denoted by Greek letters, as superscripts and subscripts. Indices corresponding three-dimensional Cartesian coordinates are denoted by Latin letters, they are always  subscripts.

If only two-dimensional tangential to the surface tensors were necessary then the well-known mathematics of differential  Riemannian geometry would be sufficient. However, Eq. (\ref{eq:intbound}) contains three-dimensional tensors on the surface rather then two-dimensional tensors. This is why the standard technique of Riemannian geometry should be modified here. Such a modification is quite natural if the frame formalism is used.

Let us introduce the frame vectors
\begin{equation}
\label{eq:framedef}
e_{i \alpha}= \partial r_i / \partial x^{\alpha}=r_{i,\alpha} \, ,
\end{equation}
where $r_i$ are Cartesian coordinates of a point on the surface. Consequently, if point moves along the surface then
\begin{equation}
(d{\bf r})^2=e_{i\alpha}e_{i\beta}dx^{\alpha}dx^{\beta} \, .
\end{equation}
So that metric tensor  of the surface coordinate system $g_{\alpha\beta}$  can be  represented as follows
\begin{equation}
\label{eq:gab}
g_{\alpha\beta}=e_{i\alpha}e_{i\beta} \, .
\end{equation}

Since generally the coordinates $x^{\alpha}$  are curvilinear, the basis set $e_{i\alpha}$ is neither orthogonal nor normalized. Therefore it is necessary to introduce  a mutual basis set $e_i^{\alpha}$, which is defined by relations
\begin{equation}
\label{eq:dab}
e_i^{\alpha}e_{i\beta}=\delta^{\alpha}_{\beta} \, ,
\end{equation}
where $\delta^{\alpha}_{\beta}$ is Kronecker delta in Greek indices. By means of the reverse metric tensor $g^{\alpha\beta}$ mutual reference vectors $e_i^{\alpha}$ can be represented in terms of $e_{i\alpha}$:
\begin{equation}
\label{eq:mutual}
e_i^{\alpha}=g^{\alpha\beta}e_{i\beta} \, .
\end{equation}

Thus, by means of  $g_{\alpha\beta}$ and $g^{\alpha\beta}$ curvilinear indices can be raised and lowered in the standard manner. Surface integrals can be represented in the form of integration over  $x^{\alpha}$ using obvious  relation: $dS=\sqrt{g}\, d^2x$, where $d^2x=dx^1dx^2$, $g={\rm det}\, g_{\alpha\beta}$. 

The key observation needed further  is that  $e_i^{\alpha}e_{j\alpha}$ is the three-dimensional projector onto the plane tangential to the surface of the body. Indeed, $e_i^{\alpha}e_{j\alpha}e_j^{\beta}e_{k\beta}=
e_i^{\alpha}\delta_{\alpha}^{\beta}e_{k\beta}=e_i^{\alpha}e_{k\alpha}$. Moreover any vector $v_j$ convolved with $e_i^{\alpha}e_{j\alpha}$ yields  a linear combination of vectors $e_i^{\alpha}$ which are tangential to the body surface.  These two facts prove our assertion. It is also  obvious that $n_in_j$ is the projector onto normal of the surface. Thus, the unit operator (Kronecker delta in Latin indices) can be represented as follows:
\begin{equation}
\label{eq:uniop}
\delta_{ij}=n_in_j + e_i^{\alpha}e_{j\alpha} \, .
\end{equation}

Eq. (\ref{eq:uniop}) allows us to convert $\delta u_{i,j}$ in such manner: $\delta u_{i,j} =\delta u_{i,l}n_ln_j + \delta u_{i,l}e_{l\alpha}e_j^{\alpha}$. From Eq. (\ref{eq:framedef}) it follows that $\delta u_{i,l}e_{l\alpha}=(\delta u_i)_{,\alpha}$, where $(\dots)_{,\alpha}$ means the partial derivative  in $x^{\alpha}$. Therefore,
\begin{equation}
\oint (\dots)\delta u_{i,j}dS =
\oint (\dots)\delta u_{i,l}n_ln_j \sqrt{g}d^2x + 
\oint (\dots) e_j^{\alpha}(\delta u_i)_{,\alpha}\sqrt{g}d^2x \, .
\end{equation}
The second term in the right hand side of this equation can be integrated by parts in straightforward manner. It  yields
\begin{equation}
\label{eq:transf}
\oint (\dots)\delta u_{i,j}dS =
\oint (\dots)\delta u_{i,l}n_ln_j dS - 
\oint\left[(\dots)e_j^{\alpha}\sqrt{g}\right]_{,\alpha}\delta u_i\,  g^{-1/2}dS \, .
\end{equation}

Identity  (\ref{eq:transf}) allows to transform the surface integral contained not independent  variations $\delta u_{i,j}$   to the surface integrals  contained $\delta u_i$ and $\delta u_{i,l}n_l$ which are independent.  Indeed, $\delta u_i$ are independent initially. As for $\delta u_{i,l}n_l\, $ it is  nothing but the gradient in the direction normal to the surface. Setting $\delta u_i$ on the surface one can continue them arbitrary along the normal to the surface. So that near  surface $\delta u_{i,l}n_l\,$ are arbitrary independent values.

Thus, by applying (\ref{eq:transf}) to (\ref{eq:intbound}) and taking into account the independence and arbitrariness of  $\delta u_i$ and $\delta u_{i,l}n_l\,$, we obtain the boundary conditions in the form
\begin{equation}
\label{eq:difb1}
\Sigma_{ij}n_j + \Theta_{ijk}n_jn_k=0 \, ,
\end{equation}
\begin{equation}
\label{eq:difbound}
\sigma_{ij}n_j+F_i -
\left(\Sigma_{ij}e_j^{\alpha}g^{1/2}\right)_{,\alpha}g^{-1/2} - 
\left(\Theta_{ijk}n_ke_j^{\alpha}g^{1/2}\right)_{,\alpha}g^{-1/2}=0 \, .
\end{equation}
Naturally, these equations are  valid only on the body surface. They are boundary conditions for differential equations of elastic equilibrium (\ref{eq:sigequil}).

Using the Leibniz rule, identity $\Theta_{ijk,\alpha}=\Theta_{ijk,l}r_{l,\alpha}$,  Eq. (\ref{eq:framedef}) and   Eq. (\ref{eq:uniop}) the last term in left hand side of (\ref{eq:difbound})  can be represented in the form  used  in the paper \cite{bib:yurk11}:
\begin{equation}
\label{eq:thetdifa}
\left(\Theta_{ijk}n_ke_j^{\alpha}g^{1/2}\right)_{,\alpha}g^{-1/2}=
\Theta_{ijk,j}n_k - \Theta_{ijk,l}n_ln_jn_k + \Theta_{ijk}\gamma_{jk} \, ,
\end{equation}
where
\begin{equation}
\label{eq:gamdef}
\gamma_{jk}=\left(n_k e_j^{\alpha}g^{1/2}\right)_{,\alpha}g^{-1/2} \, .
\end{equation}

Generally an analogous transformation of  $\left(\Sigma_{ij}e_j^{\alpha}g^{1/2}\right)_{,\alpha}g^{-1/2}$ may be  meaningless because $\Sigma_{ij}$ may be defined only on the surface rather than space. However, there are special case when  $\Sigma_{ij}$ can be represented as convolution:
\begin{equation}
\label{eq:sijldef}
\Sigma_{ij}= S_{ijl}n_l \, ,
\end{equation}
where tensor $S_{ijl}$ is well-defined in the bulk of the body. In particular, such a possibility appears  when the surface contribution to the thermodynamic potential arises due to the transformation of volume contribution by integration by parts. For such cases $\left(\Sigma_{ij}e_j^{\alpha}g^{1/2}\right)_{,\alpha}g^{-1/2}$ can be transformed absolutely similarly as 
$\left(\Theta_{ijk}n_ke_j^{\alpha}g^{1/2}\right)_{,\alpha}g^{-1/2}$. 

Equations (\ref{eq:difb1}), (\ref{eq:difbound}) with the equalities (\ref{eq:thetdifa}), (\ref{eq:gamdef}) are nothing but some  generalization of the boundary conditions obtained in \cite{bib:yurk11}. Generalization is as follows. First, we have consider the general case of the thermodynamic potential depended on the strain gradient (see previous section) rather then  a special thermodynamic potential used in \cite{bib:yurk11}. Second, we have taken into account the surface contribution to this potential, it has led to additional therm $\Sigma_{ij}n_j$ in  (\ref{eq:difb1}) and additional terms $F_i$ and $-\left(\Sigma_{ij}e_j^{\alpha}g^{1/2}\right)_{,\alpha}g^{-1/2}$ in (\ref{eq:difbound}). Except for these generalizations boundary conditions derived above coincide with ones derived in \cite{bib:yurk11}.

Note that boundary conditions  (\ref{eq:difb1}), (\ref{eq:difbound}) do satisfy the condition that in the limit case they should correspond the classical theory of elasticity. Indeed in the classical limit $\Sigma_{ij}=0$, $\Theta_{ijk}=0$. So that  (\ref{eq:difb1}) becomes an identity while (\ref{eq:difbound}) takes the classical form $\sigma_{ij}n_j + F_i=0$.

Remarkable fact   is that curvature of the body surface plays an important role in the mechanical boundary conditions derived above. To see it let us rewrite (\ref{eq:difbound}) by means (\ref{eq:thetdifa}) omitting surface contribution contained  $\Sigma_{ij}$. It turns out:
\begin{equation}
\label{eq:difboundmod}
\sigma_{ij}n_j+F_i - \Theta_{ijk,j}n_k + \Theta_{ijk,l}n_ln_jn_k - \Theta_{ijk}\gamma_{jk} 
= 0 \, .
\end{equation}
Tensor $\gamma_{jk}$ is a purely geometric construction (see also Appendix \ref{appendix0}). It does not depend on the elastic displacements or somewhat else of non-geometrical nature,  actually it  characterizes only the  geometry of the body surface. Note also that this tensor is invariant with respect to the change of the surface coordinate system. It is clear from the fact that it is a scalar convolution in Greek indices. If some part of a body surface is flat then suitable Cartesian coordinates can be chosen as surface coordinates on this part. But in such surface coordinates all values under derivative in  (\ref{eq:gamdef}) are constants. So that $\gamma_{jk}=0$ on the flat parts of the body surface. Otherwise   $\gamma_{jk} \ne 0$ on the curved parts of the body surface, it is absolutely obvious. Thus, the last term in left hand side of  (\ref{eq:difboundmod}) appears or disappears depending on whether the body surface is curved or flat. A similar conclusion can be drawn regarding the surface term  $\left(\Sigma_{ij}e_j^{\alpha}g^{1/2}\right)_{,\alpha}g^{-1/2}\,$, one should only transform it  by means of  Leibniz rule.

\section{Curvilinear coordinates and covariant formalism}\label{sec:covar}

In the previous section the mechanical boundary conditions were derived in terms of  tensor components related to the three-dimensional Cartesian coordinate system. To do this one has  introduce an additional curvilinear two-dimensional coordinate system on the surface of the body. 

One can go the other way and just introduce curvilinear coordinate system $x^{\alpha}$ in the whole space. If it is required that the  equation of the body surface in this curvilinear coordinate system has a simple form \mbox{$x^3=x^3_S\,$}, where $x^3_S$ is some constant, then the derivation of the boundary conditions can be done in a more direct way. It is essential that such a coordinate system additionally is  most convenient for the practical solution of differential equations of elastic equilibrium.

When curvilinear coordinates are used the space  remains flat. However,  to work with tensor objects at these case is  most convenient in framework  of the mathematical tool of Riemann geometry well known from the general theory of relativity. The fact that space is still flat   only leads to that the Riemann tensor is identically zero.
 
Although the above-mentioned mathematical formalism (so called covariant formalism) is well known in theoretical physics, it is not widely used in material science. This is why  we describe  its main points in this section for convenience.  It is also needed for references.

It is convenient to assume that besides curvilinear coordinate system there is  Cartesian coordinate system also. So that we can introduce frame vectors $e_{i\alpha}$ by the same equation (\ref{eq:framedef}) from the previous section. The only difference is that now Greek indices run three values 1, 2 and 3 and now $r_i$ means Cartesian coordinates of any point of space. Taking into account this comment  the equations (\ref{eq:gab}), (\ref{eq:dab}) and (\ref{eq:mutual}) from  previous section also remain valid. Contrariwise equation (\ref{eq:uniop}) now is changed into
\begin{equation}
\label{eq:deltaine}
e_i^{\alpha}e_{j\alpha} = \delta_{ij} \, .
\end{equation}

The curvilinear contravariant tensor components are defined by Cartesian tensor components as follows:
\begin{equation}
A^{\alpha_1\dots\alpha_n}=A_{i_1\dots i_n}e_{i_1}^{\alpha_1}\dots e_{i_n}^{\alpha_n} \, .
\end{equation}
The curvilinear covariant tensor components are defined analogously:
\begin{equation}
A_{\alpha_1\dots\alpha_n}=A_{i_1\dots i_n}e_{i_1\alpha_1}\dots e_{i_n\alpha_n} \, .
\end{equation}
By analogy it is  clear how are  defined mixed tensor components. By means (\ref{eq:deltaine}) it is obvious how curvilinear tensor components can be transformed back to Cartesian tensor components. Say for a tensor in covariant components this transformations looks as follows:
\begin{equation}
A_{i_1\dots i_n} = A_{\alpha_1\dots\alpha_n}e_{i_1}^{\alpha_1}\dots e_{i_n}^{\alpha_n} \, .
\end{equation} 
Tensor convolutions   are  transformed by means of (\ref{eq:dab}). For example
\begin{equation}
A_{\dots i \dots}B_{\dots i \dots}= A_{\dots i \dots}\delta_{ij} B_{\dots j \dots}= 
A_{\dots i \dots}e_i^{\alpha} e_{j\alpha}B_{\dots j \dots} \, .
\end{equation}
Thus  it is obtained automatically that only  upper and lower Greek indices can be convolved.

Gradients of tensors are transformed in a special way. Since the frame  vectors are not constant they can not be  moved  through the derivative in straightforward manner, additional terms appear during this motion. For example gradient of  vector is transformed as follows:
\begin{equation}
\label{aij}
A_{i,j}e_{i\alpha}e_{j\beta}=A_{\alpha,\beta} - A_{\gamma}\Gamma^{\gamma}_{\alpha\beta} \, ,
\end{equation}
where
\begin{equation}
\label{gamm0}
\Gamma^{\gamma}_{\alpha\beta}  =    e_i^{\gamma} e_{i\alpha,j}e_{j\beta} =   
e_i^{\gamma} e_{i\alpha,\beta}    \,  .
\end{equation}
The expression in the right hand side of   (\ref{aij})  is called the covariant derivative and is denoted by index separated by a semicolon:
\begin{equation}
A_{\alpha;\beta} =  A_{\alpha,\beta} - A_{\gamma}\Gamma^{\gamma}_{\alpha\beta} \, .
\end{equation} 
Set of values $\Gamma^{\gamma}_{\alpha\beta}$ is called the Christoffel symbols. They can be expressed not only in terms of frame vectors (\ref{gamm0}) but also in terms of metric tensor
\begin{equation}
\label{gg1}
\Gamma^{\gamma}_{\alpha\beta}=\frac{1}{2} g^{\gamma\delta}
(g_{\delta\alpha,\beta}+g_{\delta\beta,\alpha}-g_{\alpha\beta,\delta}) \,  .
\end{equation}
It is easy to see that this  Christoffel symbols are symmetric in the lower indices. There is also useful equation
\begin{equation}
\label{eq:divgamma}
\Gamma^{\beta}_{\alpha\beta} =\frac{g_{,\alpha}}{2g}=(\ln\sqrt{g})_{,\alpha} \, 
\end{equation}
which can be obtained if one differentiate $\ln g = {\rm Sp} \ln g_{\alpha\beta}\,$, $g={\rm det} g_{\alpha\beta}$ hereafter.

The covariant derivative of contravariant vector should be defined in another manner. The reason is that frame vector $e_i^{\alpha}$ should be moved through the derivative  rather then $e_{i\alpha}\,$. Note that $e_i^{\alpha}e_{i\beta}$ is a constant (see equation (\ref{eq:dab})), so that it partial derivative is zero. Therefore \mbox{$e_{i,\gamma}^{\alpha}e_{i\beta}=-e_i^{\alpha}e_{i\beta,\gamma}\,$} and from this identity and simple algebra it turns out:
\begin{equation}
A_{i,j}e_i^{\alpha}e_{j\beta}=
A^{\alpha}_{\phantom{\mu};\beta}=
A^{\alpha}_{\phantom{\mu},\beta} + A^{\gamma}\Gamma^{\alpha}_{\gamma\beta}
\, .
\end{equation}

From the above it is quite clear how to determine the covariant derivative of higher rank tensors . The terms containing the Christoffel symbols (so-called $\Gamma$-terms) appear  due to frame  vectors motion   through the derivative. If the tensor is of the highest rank then several frame vectors should be moved  through  derivative. Obviously several $\Gamma$-terms should appear in this way. For example covariant derivative of two-rank tensor with lower indices  is
\begin{equation}
A_{\alpha\beta;\gamma} = A_{\alpha\beta,\gamma} -
A_{\delta\beta}\Gamma^{\delta}_{\alpha\gamma} -
A_{\alpha\delta}\Gamma^{\delta}_{\beta\gamma} \, .
\end{equation}
What happens in the other cases  is clear by analogy. Covariant derivative of a scalar is determined as usual partial derivative.

It is easy to show that the covariant derivative of the product obeys  the Leibniz rule
\begin{equation}
\label{leibnitz}
(A_{\dots}B_{\dots})_{;\alpha} = A_{\dots;\alpha}B_{\dots}+A_{\dots}B_{\dots;\alpha} \,  .
\end{equation}
It can be also proved in straightforward manner that the covariant derivative of frame vectors is identically zero. Since metric tensor can be represented as the convolution (in Cartesian indices) of frame vectors, it covariant derivative is identically zero also. It is also worth  that if some tensor is constant in Cartesian components (say it is a material tensor of homogeneous media) then in curvilinear components  it covariant (but not usual partial) derivatives are zero.

In the general case of curved space covariant derivatives do not commute, their commutator is proportional to the Riemann curvature tensor. However, here we are interested only in the case of curvilinear coordinates in flat space. In the flat space Riemann tensor is identically zero. Therefore,  the covariant derivatives commute.

To complete this brief overview  of covariant tensor analysis it remains to say few words about integration. As for the volume integration  everything is quite simple. It is easy to see that the Jacobian of the transformation from Cartesian coordinates to curvilinear ones is the determinant of  $e_{i\alpha}\,$. However, it is more convenient and  generally accepted to express this  Jacobian in terms  of  metric tensor. It is clear from  (\ref{eq:gab}) that the square of the Jacobian equal to $g$. So that
\begin{equation}
dV=\sqrt{g}\,d^3x \, ,
\end{equation}
where $d^3x=dx^1dx^2dx^3\,$. This formula is enough for volume integration, at least in the part that we need here.

As for the integration over the surface, the situation is somewhat more complicated.  Here we restrict ourselves to the case when the equation of the surface has a simple form $x^3={\rm const}\,$. The more general case we do not need.

If the equation of the surface has the simple form mentioned above then the integration over the surface reduces to integration over $x^1$ and $x^2$. Let us introduce additional notation: Greek indices in parentheses have the values 1 and 2 only. In this notation it turns out
\begin{equation}
dS=\sqrt{\bf g}\,d^2x \, ,
\end{equation}
where $d^2x=dx^1dx^2\,$, ${\bf g}$ is determinant of  $2 \times 2$ matrix $g_{(\alpha)(\beta)}\,$. Note that according to the known representation of the inverse matrix by algebraic adjuncts
\begin{equation}
\label{eq:bgg33g}
{\bf g}=g^{33}g \, .
\end{equation}

Surface integrals often contain an unit vector normal to the surface. So that we also need the expression for such a vector. It is clear geometrically that in the considered special coordinate system the vectors $e_{i 1}$ and $e_{i 2}$ are tangential  to the surface. Therefore in accordance with (\ref{eq:dab}) the vector $e^3_i$ is normal to the surface. By making the normalization of this vector, we obtain the unit  vector normal to  the surface in Cartesian components:
\begin{equation}
n_i=\frac{e_i^3}{\sqrt{e^3_ie^3_i}}=\frac{e_i^3}{\sqrt{g^{33}}} \, .
\end{equation}
Conversion of this vector to the curvilinear components by means of general equations  yields the following:
\begin{equation}
\label{eq:covarn}
n_{\alpha}=n_ie_{i\alpha}=\frac{e_i^3e_{i\alpha}}{\sqrt{g^{33}}}=
\frac{\delta^{3}_{\alpha}}{\sqrt{g^{33}}} 
\, ,
\end{equation}
\begin{equation}
\label{eq:contravarn}
n^{\alpha}=g^{\alpha\beta}n_{\beta}=
\frac{g^{\alpha 3}}{\sqrt{g^{33}}} 
\, .
\end{equation}

The above  brief consideration of the covariant formalism   is sufficient for the purposes of subsequent exposition. For more details one should refer to any textbook on general relativity.

\section{Boundary conditions within covariant formalism}\label{sec:covarbound}

In framework of mathematics briefly described in the previous section equations (\ref{eq:sigdef}) -- (\ref{eq:intbound}) take the form
\begin{equation}
\sigma^{\alpha\beta}=T^{\alpha\beta} - \Theta^{\alpha\beta\gamma}_{\phantom{\alpha\beta\gamma} ; \gamma}  \, ,
\end{equation}
\begin{equation}
\sigma^{\alpha\beta}_{\phantom{\alpha\beta} ; \beta}=0 \, ,
\end{equation}
\begin{equation}
\label{eq:covintbound}
\oint \left( \sigma^{\alpha\beta}n_{\beta}  + F^{\alpha}\right)\delta u_{\alpha}\sqrt{\bf g} \, d^2x +
\oint \left(\Sigma^{\alpha\beta} + \Theta^{\alpha\beta\gamma}n_{\gamma}\right) 
\delta u_{\alpha ; \beta}\sqrt{\bf g} \, d^2x = 0 \, .
\end{equation} 
One should stress that equation of the body surface of  the form $x^3=x^3_S={\rm const}$ is implied here, otherwise the last equation is not valid. Should  be noted also that the  differentiation of scalar in covariant tensor components yields  contravariant tensor components. So that
\begin{equation}
T^{\alpha\beta} = \frac{\partial {\cal H}_B}{\partial u_{\alpha ; \beta}} \, ,
\end{equation}
\begin{equation}
\Theta^{\alpha\beta\gamma} = \frac{\partial {\cal H}_B}{\partial u_{\alpha ; \beta ; \gamma}} \, ,
\end{equation}
\begin{equation}
F^{\alpha}=\frac{\partial {\cal H}_S}{\partial u_{\alpha}} \, ,
\end{equation}
\begin{equation}
\Sigma^{\alpha\beta}=\frac{\partial {\cal H}_S}{\partial u_{\alpha ; \beta}} \, .
\end{equation}

Transformation of the second integral in left hand side of (\ref{eq:covintbound}) can be provided quite directly. One should represent $\delta u_{\alpha ; \beta}$ in terms of partial derivative and $\Gamma$-term
\begin{equation}
\delta u_{\alpha ; \beta}=\delta u_{\alpha , \beta} - 
\delta u_{\gamma}\Gamma^{\gamma}_{\alpha\beta} \, 
\end{equation}
and  separate the case of $\beta=3$ in the terms contained $\delta u_{\alpha , \beta}\,$ as follows:
\begin{equation}
(\dots^{\alpha\beta})\delta u_{\alpha , \beta} = (\dots^{\alpha 3})\delta u_{\alpha , 3} + (\dots^{\alpha(\beta)})\delta u_{\alpha , (\beta)} \, .
\end{equation}
Remember that index in  parentheses means the index which values are 1 and 2 only. Integral contained $\delta u_{\alpha , (\beta)}$ can be integrated by parts without any tricks. In this way it turns out:
\begin{equation}
\label{eq:covintbount}
\oint  L^{\alpha} \delta u_{\alpha} \sqrt{\bf g} d^2x + 
\oint \left(\Sigma^{\alpha 3} + \Theta^{\alpha 3 \gamma}n_{\gamma}\right)
\delta u_{\alpha , 3} \sqrt{\bf g} d^2 x =0 \, ,
\end{equation}
where
\begin{equation}
\label{eq:la}
\begin{array}{l}
\displaystyle
L^{\alpha} = \sigma^{\alpha\beta}n_{\beta} + F^{\alpha} - 
\Sigma^{\delta\varepsilon}\Gamma^{\alpha}_{\delta\varepsilon}  -
\left( \Sigma^{\alpha(\beta)}  {\bf g}^{1/2} \right)_{, (\beta)} {\bf g}^{-1/2} -
\\ \\
\displaystyle
\Theta^{\delta\varepsilon\gamma}n_{\gamma}\Gamma^{\alpha}_{\delta\varepsilon} - 
\left( \Theta^{\alpha(\beta)\gamma}n_{\gamma}
{\bf g}^{1/2} \right)_{, (\beta)} {\bf g}^{-1/2} \, .
\end{array}
\end{equation}
Note that since  (\ref{eq:bgg33g}) and  (\ref{eq:covarn}) $L^{\alpha}$  can be rewritten using $g$ rather then ${\bf g}$ in last term:
\begin{equation}
\label{eq:la2}
\begin{array}{l}
\displaystyle
L^{\alpha} = \sigma^{\alpha\beta}n_{\beta} + F^{\alpha} - 
\Sigma^{\delta\varepsilon}\Gamma^{\alpha}_{\delta\varepsilon}  -
\left( \Sigma^{\alpha(\beta)}  {\bf g}^{1/2} \right)_{, (\beta)} {\bf g}^{-1/2} -
\\ \\
\displaystyle
\Theta^{\delta\varepsilon\gamma}n_{\gamma}\Gamma^{\alpha}_{\delta\varepsilon} - 
\left( \Theta^{\alpha(\beta)\gamma}
g^{1/2} \right)_{, (\beta)}  g^{-1/2}n_{\gamma} \, .
\end{array}
\end{equation}

From (\ref{eq:covintbount}) the boundary conditions  are obvious and have the following form
\begin{equation}
\label{eq:covarbound1}
\Sigma^{\alpha 3} + \Theta^{\alpha 3 \gamma}n_{\gamma} = 0 \, ,
\end{equation}
\begin{equation}
\label{eq:covarbound2}
L^{\alpha} = 0 \, .
\end{equation}
Note that since $n_{\alpha} \sim \delta^3_{\alpha}$ (see (\ref{eq:covarn})) (\ref{eq:covarbound1}) can be rewritten in the form
\begin{equation}
\label{eq:covarmondnn}
\Sigma^{\alpha \beta}n_{\beta} + 
\Theta^{\alpha \beta \gamma}n_{\beta}n_{\gamma} = 0 \, 
\end{equation}
which corresponds to  (\ref{eq:difb1}). As regards (\ref{eq:covarbound2}) with (\ref{eq:la}) or (\ref{eq:la2})  it looks  different then (\ref{eq:difbound}).  Nevertheless it can be proved that   (\ref{eq:covarbound2}) is equivalent to (\ref{eq:difbound}). The proof follows below.

Before the description of the  proof itself we should discuss  some geometrical matter. The curvilinear coordinate system used in this section is such that equation of body surface is $x^3={\rm const}$. Thus, first two coordinates $x^{(\alpha)}$  of this coordinate system  may be well  used as surface coordinates system used in section~\ref{sec:cartbound}. Nevertheless, in this case  one  one should distinguish some objects depending on whether $x^{(\alpha)}$ are used as first two coordinates of spatial coordinate system or as the surface coordinate system in manner of section~\ref{sec:cartbound}.  For such a distinction here  we add additional tilde to objects related to the surface coordinate system.

Not that $\tilde{e}_{i(\alpha)}= e_{i(\alpha)}$ but   $\tilde{e}_i^{(\alpha)} \ne e_i^{(\alpha)}$. This is consequence of the fact that $\tilde{g}^{(\alpha)(\beta)} \ne g^{(\alpha)(\beta)}$.  By means of simple geometrical reasoning one can see that
\begin{equation}
\label{eq:tildee}
\tilde{e}_i^{(\alpha)}=e_i^{(\alpha)} - \frac{ g^{(\alpha)3}}{g^{33}} e_i^3 \, .
\end{equation}
Note also that
\begin{equation}
\label{eq:tildeg}
\tilde{g}={\bf g} \, .
\end{equation}

Now one can turn to the proof itself. Let us denote last two terms in  (\ref{eq:difbound}) as $\phi_i\,$. By means of (\ref{eq:tildee}) and (\ref{eq:tildeg}) we can rewrite it in terms of 3-dimensional coordinates:
\begin{equation}
\label{eq:phii}
\begin{array}{l}
\displaystyle
\phi_i = - \left(\Sigma_{ij}e_j^{(\alpha)} {\bf g}^{1/2}\right)_{, (\alpha)} {\bf g}^{-1/2} +
\left(\Sigma_{ij}\frac{ g^{(\alpha)3}}{g^{33}} e_j^3 {\bf g}^{1/2}\right)_{, (\alpha)}{\bf g}^{-1/2} - \\ \\
\displaystyle
\left(\Theta_{ijk}n_ke_j^{(\alpha)} {\bf g}^{1/2}\right)_{, (\alpha)} {\bf g}^{-1/2} +
\left(\Theta_{ijk}n_k\frac{ g^{(\alpha)3}}{g^{33}} e_j^3 {\bf g}^{1/2}\right)_{, (\alpha)}{\bf g}^{-1/2}
\, .
\end{array}
\end{equation}
Second and fourth terms in right hand side of this equation can be  combined into one term
\begin{equation}
\label{eq:zeroterms}
\left[\left(\Sigma_{ij}e_j^3+\Theta_{ijk}n_ke_j^3\right)
 \frac{ g^{(\alpha)3}}{g^{33}}  {\bf g}^{1/2}\right]_{, (\alpha)}{\bf g}^{-1/2} \, .
\end{equation}
Since $e^3_i \sim n_i$ the expression in parentheses is zero due to (\ref{eq:difb1}). Moreover, since it is zero everywhere on the surface (i.e. for every $x^{(\alpha)}$),  its derivative in $x^{(\alpha)}$ is zero also. So that (\ref{eq:zeroterms}) is zero i.e.  second and fourth terms in  (\ref{eq:phii}) cancel each other  out completely. Thus,
\begin{equation}
\label{eq:phii2}
\phi_i = - \left(\Sigma_{ij}e_j^{(\beta)} {\bf g}^{1/2}\right)_{, (\beta)} {\bf g}^{-1/2} -
\left(\Theta_{ijk}n_ke_j^{(\beta)} {\bf g}^{1/2}\right)_{, (\beta)} {\bf g}^{-1/2} 
\, .
\end{equation}
Further one can easily  calculate contravariant representation of vector $\phi_i$. It turns out:
\begin{equation}
\label{eq:phialpha}
\begin{array}{l}
\displaystyle
\phi^{\alpha}=e^{\alpha}_{i , (\beta)}e^{(\beta)}_j\left(\Sigma_{ij}+\Theta_{ijk}n_k
\right) - \\ \\
\displaystyle
\left(\Sigma^{\alpha(\beta)}{\bf g}^{1/2}\right)_{, (\beta)} {\bf g}^{-1/2} -
\left(\Theta^{\alpha(\beta)\gamma}n_{\gamma}{\bf g}^{1/2}\right)_{, (\beta)} {\bf g}^{-1/2}
 \, .
 \end{array}
\end{equation}
In the first term of right hand side of this equation range of the index $\beta$  can be expanded to complete set 1,2,3.   This is consequence of (\ref{eq:difb1}) and the fact that   $e^3_i \sim n_i$.  Further one should note that
\begin{equation}
e^{\alpha}_{i , \beta}e^{\beta}_j=
e^{\alpha}_{l , \beta}\delta_{li}e^{\beta}_j =
e^{\alpha}_{l , \beta}e_{l\delta}e^{\delta}_ie^{\beta}_j=
-  \Gamma^{\alpha}_{\beta\delta} e^{\delta}_i e^{\beta}_j \, .
\end{equation} 
With this identity equation  (\ref{eq:phialpha}) takes the following form 
\begin{equation}
\label{eq:phialpha2}
\begin{array}{l}
\displaystyle
\phi^{\alpha}=  -\Gamma^{\alpha}_{\delta(\beta)}\left(\Sigma^{\delta(\beta)}+
\Theta^{\delta(\beta)\gamma}n_{\gamma} \right) - \\ \\
\displaystyle
\left(\Sigma^{\alpha(\beta)}{\bf g}^{1/2}\right)_{, (\beta)} {\bf g}^{-1/2} -
\left(\Theta^{\alpha(\beta)\gamma}n_{\gamma}{\bf g}^{1/2}\right)_{, (\beta)} {\bf g}^{-1/2}
 \, .
 \end{array}
\end{equation}
Note that range of index $\beta$ in first term of right hand side of this equation can be extended to complete set 1,2,3. This is a consequence of the equation (\ref{eq:covarbound1}). With this extension further elementary  transformations yield exactly (\ref{eq:covarbound2}). Proof  is completed.

Also it is worth to transform obtained boundary conditions (\ref{eq:covarbound1}) and (\ref{eq:covarbound2}) to the forms used  in papers \cite{bib:yurk13,bib:yurk14}. In this papers only case of $F^{\alpha}=0$ and $\Sigma^{\alpha\beta}=0$  considered, the same is assumed  while the transformation.  So that by means of (\ref{eq:covarn}) equation  (\ref{eq:covarbound1}) can be rewritten as
\begin{equation}
\label{eq:eq:covarbound3}
\Theta^{\alpha 3 3} = 0 \, .
\end{equation}
By means of Leibniz rule, equation (\ref{eq:divgamma}) and representation of partial derivative in terms of covariant derivative last  term in right hand side of (\ref{eq:la2}) can be transformed as follows
\begin{equation}
\begin{array}{l}
\displaystyle
\left( \Theta^{\alpha(\beta)\gamma}g^{1/2} \right)_{, (\beta)}  g^{-1/2}n_{\gamma}=
\Theta^{\alpha\beta\gamma}_{\phantom{\alpha(\beta) \gamma} ; \beta}n_{\gamma} -
\Theta^{\delta\beta\gamma}\Gamma^{\alpha}_{\delta\beta}n_{\gamma} - \\ \\
\displaystyle
\Theta^{\alpha\beta\delta}\Gamma^{\gamma}_{\delta\beta}n_{\gamma} - 
\left(\Theta^{\alpha 3 \gamma}g^{1/2}\right)_{,3}g^{-1/2}n_{\gamma} \, .
\end{array}
\end{equation} 
With this identity (\ref{eq:covarbound2}) takes the form
\begin{equation}
\label{eq:eq:covarbound4}
\sigma^{\alpha\beta}n_{\beta} -
\Theta^{\alpha\beta\gamma}_{\phantom{\alpha(\beta) \gamma} ; \beta}n_{\gamma} +
\Theta^{\alpha\beta\delta}\Gamma^{\gamma}_{\delta\beta}n_{\gamma} + 
\left(\Theta^{\alpha 3 \gamma}g^{1/2}\right)_{,3}g^{-1/2}n_{\gamma} =0 \, .
\end{equation}
This  coincides with (12) from \cite{bib:yurk13}. By means (\ref{eq:covarn}) and (\ref{eq:eq:covarbound3}) it can be also rewritten in following form
 \begin{equation}
\label{eq:eq:covarbound5}
\sigma^{\alpha 3} - 
\Theta^{\alpha\beta 3}_{\phantom{\alpha(\beta) \gamma} ; \beta} +
\Theta^{\alpha(\beta\delta)}\Gamma^{3}_{(\delta\beta)} +
\Theta^{\alpha 3 3}_{\phantom{\alpha 3 3} , 3} =0 \, ,
\end{equation}
where one pair of parentheses enclose several indexes mean that these indices are not equal 3 simultaneously. Boundary conditions (\ref{eq:eq:covarbound3}) and (\ref{eq:eq:covarbound5}) are exactly the ones  used in \mbox{\cite{bib:yurk14}}.

\section{Conclusion}

It is described above how one can generalize classical theory of elastic continuum to a case when thermodynamic potential depends on  the strain and the strain gradients. In this case, in fact, one gets such a type of classical (non-quantum) field theory were function which determinates  field equations (``Lagrangian'') depends on the first and second derivatives of the initial field, particularly the elastic displacement field $u_i\,$. Certainly in the static case ``Lagrangian'' is nothing but the thermodynamic potential.   

At the same time it should be noted that in essence we did not use the fact that the field $u_i$ is  a field of elastic displacements. All this is also true in cases where the vector field $u_i$ has another physical nature if only ``Lagrangian'' has the form that we has expected. It is not a significant problem to add also  a ``mass term''  (in terms of the quantum field theory) $a_{ij}u_iu_j$ where $a_{ij}$ is some constant tensor. It will not change much, only differential equations will be slightly (and obviously) modified while the boundary conditions will remain the same. Such opportunities should be kept in mind, but practically considered theory is needed to describe namely the elastic fields, and mostly in relation to  the flexoelectric effect.

It is shown in this paper  that in theories of this type, particularly in generalized theory of elasticity, differential equations can be derived in quite conventional manner. The only difference with respect to ordinal case is  that while  varying  one should  integrate by parts twice. As for the derivation  of the boundary conditions for these differential  equations, the situation is more complicated. We have considered two ways by which such derivation  can be done. The first way uses a mathematical trick based on the use of additional surface coordinate system. The second way is more direct, but it involves the use of a covariant tensor analysis and a curvilinear coordinate system of a special class. In the coordinate systems of this class equation of  boundary surface should have the form $x^3=x^3_S$ where $x^3_S$ is some constant. For a complex surface shape it may be difficult to find a single coordinate system of this type, but in this case  it is  possible to divide the surface to  such parts, for which the  finding of the coordinate system of the desired type is not very difficult. It is, however, necessary to ensure that these parts are sewed sufficiently smoothly.

It should be noted also that in the considered case  the differential equations and especially  boundary conditions   are very complicated. With regard to the calculation of the flexoelectric deformations corresponding boundary problem was solved for a homogeneously polarized ball of an isotropic material \cite{bib:yurk13}. Even in this simple geometry one has to use a representation of the radial functions in the form of power series. To solve  more complex boundary problems even more difficult. But for the calculation of specifically flexoelectric  deformations it was proposed an approximate method \cite{bib:yurk14} which radically simplifies the problem. Corresponding theory is considered in detail in another paper \cite{bib:yurk15}.

\section*{Acknowledgements}

A.K.~Tagantsev is acknowledged for reading the manuscript.

\begin{appendices}

\section{Relation between tensor $\gamma_{ij}$ and Gauss second fundamental form}\label{appendix0}

By means of  Leibniz rule equation (\ref{eq:gamdef}), which defines the tensor $\gamma_{ij}\,$, can be rewritten as follows: 
\begin{equation}
\label{eq:gij-ini}
\gamma_{ij}=n_{j,\alpha}e^{\alpha}_i +
n_je^{\alpha}_{i,\alpha} + 
n_je^{\alpha}_i\frac{1}{2g}g_{,\alpha} \, .
\end{equation}
It can been shown (see  Section \ref{sec:covar}) that
\begin{equation}
g_{,\alpha}=g g^{\beta\gamma}g_{\beta\gamma , \alpha} \, .
\end{equation}
Representing metric tensor and reverse metric tensor in terms of surface frame vectors, we find:
\begin{equation}
g_{,\alpha}=2ge^{\beta}_le^{\gamma}_le_{k\gamma}e_{k\beta,\alpha} \, .
\end{equation}
This equation can be essentially simplified. By means of the relation $e^{\gamma}_le_{k\gamma}=\delta_{kl}-n_ln_k\,$, it turns out
\begin{equation}
g_{,\alpha}=2g\left(e^{\beta}_ke_{k\beta,\alpha} -
e^{\beta}_le_{k\beta,\alpha}n_ln_k\right) \, . 
\end{equation}
Any surface frame vector is orthogonal to  vector $n_l$ normal to the surface. Thus, second term in   parentheses is identically zero and it turns out:
\begin{equation}
g_{,\alpha}=2g e^{\beta}_ke_{k\beta,\alpha}  \, . 
\end{equation}
Thus,  (\ref{eq:gij-ini}) can be rewritten as follows:
\begin{equation}
\label{eq:gamij0}
\gamma_{ij}=n_{j,\alpha}e^{\alpha}_i +
n_je^{\alpha}_{i,\alpha} + 
n_je^{\alpha}_i  e^{\beta}_ke_{k\beta,\alpha}  \, .
\end{equation}

Second term in right hand side of (\ref{eq:gamij0}) can be transformed in such manner that its appears the term  cancelling out  third term. First, we transform it as follows:
\begin{equation}
n_je^{\alpha}_{i,\alpha} = n_j\delta_{ik}e^{\alpha}_{k,\alpha}=
n_je^{\beta}_ie_{k\beta}e^{\alpha}_{k,\alpha} +
n_in_jn_ke^{\alpha}_{k,\alpha} \, .
\end{equation}
From the fact that $e_{k\beta}e^{\alpha}_{k\beta}=\delta^{\alpha}_{\beta}\,$ does not depend on $x^{\alpha}\,$, and obvious relation $e_{k\beta,\alpha}= e_{k\alpha,\beta}\,$,
it follows that $e_{k\beta}e^{\alpha}_{k,\alpha}= - e_{k\alpha,\beta}e^{\alpha}_k \,$. Thus, it turns out:
\begin{equation}
n_je^{\alpha}_{i,\alpha} = 
n_in_jn_ke^{\alpha}_{k,\alpha} -
n_je^{\alpha}_ie_{k\beta,\alpha}e^{\beta}_k \, .
\end{equation}
Last term in right hand side of this equation obviously cancels out third term in right hand side of  (\ref{eq:gamij0}). Thus,
\begin{equation}
\label{eq:gamij1}
\gamma_{ij}=n_{j,\alpha}e^{\alpha}_i +
n_in_jn_ke^{\alpha}_{k,\alpha}
 \, .
\end{equation}

First term in right hand side of (\ref{eq:gamij1}) can be related to Gauss second fundamental form $b_{\alpha\beta}\,$, which is defined  as follows:
\begin{equation}
b_{\alpha\beta} = n_ke_{k\beta,\alpha} \, .
\end{equation}
Since $n_ke_{k\beta}=0\,$, $n_{k,\alpha}e_{k\beta} = - n_ke_{k\beta,\alpha} \,$. Convolving this equation with  $e^{\beta}_j\,$ and taking into account that \mbox{$e_{k\beta}e^{\beta}_j=\delta_{jk}-n_jn_k\,$}, we find:
\begin{equation}
n_{j,\alpha} - n_{k,\alpha}n_jn_k =  - n_ke_{k\beta,\alpha}e^{\beta}_j \, .
\end{equation} 
From $n_kn_k=1\,$, it follows that $n_{k,\alpha}n_k=0\,$. Thus,
\begin{equation}
\label{eq:Veingarten}
n_{j,\alpha}  =   - n_ke_{k\beta,\alpha}e^{\beta}_j = - b_{\alpha\beta}  e^{\beta}_j\, ,
\end{equation} 
In the theory of surfaces this equation is known as Weingarten equation. By means of it we find:
\begin{equation}
n_{j,\alpha}e^{\alpha}_i = - B_{ij} \, ,
\end{equation}
where
\begin{equation}
\label{eq:bij}
B_{ij} =  b_{\alpha\beta}e_i^{\alpha}e_j^{\beta} \, .
\end{equation}
It should be emphasized that 3-dimensional tensor $B_{ij}$ is not a tensor of second fundamental form $b_{\alpha\beta}\,$ itself, although it  is related to $b_{\alpha\beta}\,$.

Second term in right hand side of  (\ref{eq:gamij1}) can be also represented in terms of $B_{ij}\,$. Obviously $B_{ii}=b_{\alpha\beta}g^{\alpha\beta}\,$. Next,
\begin{equation}
n_in_jn_ke^{\alpha}_{k,\alpha}=
n_in_jn_kg^{\alpha\beta}_{\phantom{\alpha\beta},\alpha}e^{\beta}_k +
n_in_jn_kg^{\alpha\beta}e_{k\beta,\alpha}\, .
\end{equation}
First term in right hand side of this equation is zero identically, because  $n_ke^{\beta}_k=0\,$. It remains only the term, which contains $n_ke_{k\beta,\alpha}=b_{\alpha\beta}\,$. This eventually yields the trace of the tensor $B_{ij}\,$. Finally it turns out:
\begin{equation}
\gamma_{ij}=n_in_jB_{kk} - B_{ij} \, . 
\end{equation}

\end{appendices}


\begin{thebibliography}{99}

\bibitem{bib:tagyud13}
P.V.~Yudin and A.K.~Tagantsev, Fundamentals of flexoelectricity in solids, \emph{Nanotechnology}. {\bf 24}, 432001--432036 (2013).

\bibitem{bib:zubko13}
P.~Zubko, G.~Catalan, A.K.~Tagantsev, Flexoelectric effect in solids, \emph{Ann.~Rev.~Mater.~Res.} {\bf 43}, 387---421 (2013). 

%%%%%%%%%%%%%%%%%%%%%%%%%%%%

\bibitem{bib:Toupin62}
R.~Toupin, Elastic materials with couple-stresses. \emph{Archive for Rational Mechanics and Analysis} {\bf 11}(1), 385--414 (1962).

\bibitem{bib:Toupin64}
R.~Toupin, Theories of elasticity with couple-stress. \emph{Archive for Rational Mechanics and Analysis} {\bf 17}(2), 85--112 (1964).

\bibitem{bib:Mindlin_Eshel_68}
R.D.~Mindlin, N.N.~Eshel, On first strain-gradient theories in linear elasticity. \emph{Int. J. Solid Structures} {\bf 4}, 109--124 (1968).

\bibitem{bib:Germain73}
P.~Germain, The method of virtual power in continuum mechanics. \mbox{Part 2:} Microstructure, \emph{SIAM Journal on Applied Mathematics} {\bf 25}(3), 556--575 (1973).

\bibitem{bib:shul10}
Hu ShuLing and Shen ShengPing, Variational principles and governing equations in nano-dielectrics with the flexoelectric effect, \emph{Sci.~China~Phys.~Mech.~Astron.} {\bf 53}(8), 1497--1504 (2010). 


%%%%%%%%%%%%%%%%%%%%%%%%%%%%


\bibitem{bib:inden81}
V.~L. Indenbom, E.~B. Loginov and M.~A. Osipov, The flexoelectric effect and the structure of crystals,      \emph{Kristallografiya}. {\bf 26}(6), 1157--1162 (1981).

\bibitem{bib:MSS06}
R.~Maranganti, N.D.~Sharma, P.~Sharma,  Electromechanical coupling in nonpiezoelectric materials due to nanoscale nonlocal size effects: Green's function solutions and embedded inclusions, \emph{Phys.~Rev.~B} {\bf 74}, 014110 (2006). 

\bibitem{bib:SMS07}
N.D.~Sharma, R.~Maranganti,  P.~Sharma,  On the possibility of piezoelectric nanocomposites without using piezoelectric materials, \emph{J.~Mech.~Phys.~Solids} {\bf 55}, 2328 -- 2350 (2008). 

\bibitem{bib:MSC08}
M.S.~Majdoub, P.~Sharma, T.~Cagin, Enhanced size-dependent piezoelectricity and elasticity in nanostructures due to the flexoelectric effect, \emph{Phys.~Rev.~B} {\bf 77}, 125424 (2008). 

\bibitem{bib:yurk11}
A.~S. Yurkov, Elastic boundary conditions in the presence of the flexoelectric effect, \emph{JETP~Letters}. {\bf 94}(6), 455--458 (2011).

\bibitem{bib:yurk13}
A.~S. Yurkov, Flexoelectric deformation of a homogeneously polarized ball, \emph{	arXiv:1304.1868 [cond-mat.mtrl-sci]}. (2013).

\bibitem{bib:yurk14}
A.~S. Yurkov, On the flexoelectric deformations of finite size bodies, \emph{JETP Letters}. {\bf 99}, 214--218 (2014). See also \emph{arXiv:1305.3874v3 [cond-mat.mtrl-sci]}. (2013).

\bibitem{bib:yurk15}
A.~S. Yurkov, On the Continual Theory of Flexoelectric Deformations, \emph{arXiv:1501.04300 [cond-mat.mtrl-sci]}. (2015)
   
\end{thebibliography}
\end{document}